# Optimization of quantum well number of AlGaN/AlGaN deep-ultraviolet light-emitting diodes

G. R. Suwito[1]*, Y.-H. Shih[2], S.-W. H. Chen[3], Z.-H. Zhang[4], and H.-C. Kuo[3]

[1]*Department of Applied Physics, Nagoya University, Nagoya 464-8603, Japan*
[2]*Department of Photonics, National Cheng Kung University, Tainan 70101, Taiwan*
[3]*Institute of Electro-Optical Engineering, National Chiao Tung University, Hsinchu 300, Taiwan*
[4]*Institute of Micro-Nano Photoelectron and Electromagnetic Technology Innovation, Hebei University of Technology, Tianjin 300401, China*

*gsuwito@alice.xtal.nagoya-u.ac.jp

**Abstract**

In this work, performance and characteristics of AlGaN/AlGaN deep-ultraviolet light-emitting diodes (DUV LEDs) with varied number of quantum-well (QW) are investigated numerically. From our simulation, 1-QW structure give the best performance at low injection current. However, at higher injection current, 2-QWs structure give the largest power output due to its higher total radiative recombination rate and internal quantum efficiency (IQE) compared to other structures. The 2-QWs structure also has less serious efficiency droop at high current than 1-QW, which makes it an optimum structure for high-power LEDs.

High-efficiency AlGaN-based deep ultraviolet light-emitting diodes (DUV LEDs) have attracted great interests in recent years due to their wide-range potential applications including purification, biochemistry and medicine.[1] Nonetheless, the present performance of AlGaN-based DUV LEDs is still far from satisfactory, though various attempts have been made.[2,3] Some challenges such as high threading dislocation density, low illumination efficiency, and low hole activation of Mg-doping limit the output performance of DUV LEDs.[4-6]

In order to optimize the illumination efficiency of UV LEDs, the internal quantum efficiency (IQE) of the active region need to be enhanced.[7] Numerous works have been made to improve IQE of UV LEDs such as elevation of crystalline quality[8,9] and active region structural optimization.[10-12] The choice of the number of quantum-well (QW) is a very important aspect in structure optimization of LEDs.[13] There have been some efforts in understanding the effect of QW number, such as the quality of the crystal degraded when the QW number increased.[14-16] It has been demonstrated numerically in AlInGaN/AlInGaN UV LEDs, that 3-QWs structure give the best performance.[17] Whereas in InGaN/GaN LEDs, the single quantum well (1-QW) structure give better performance compared to 5-QWs structure due to the non-uniformity carrier distribution across the active region.[13]

Moreover, the studies that has been conducted so far show that the correlation between QW numbers and performance of the LEDs is not simply linear. It indicates that there should be an optimized number of QW to achieve more efficient structure of nitride-based DUV LEDs. In this study, the characteristics of AlGaN/AlGaN DUV LEDs with various QW numbers are numerically investigated using a self-consistent simulation program APSYS (abbreviation of Advanced Physical Models of Semiconductor Devices).[18]

The Crosslight APSYS simulation program is a 2-D simulation which solves the Poisson's equation, photon rate equation, current continuity equations, and scalar wave equation, to give optical and electrical properties of the semiconductor devices, in particular the LEDs. APSYS also includes transport model which consists of carrier drift diffusion in the devices, carrier capture/escape, and direct flying over across QWs. APSYS employs the $6 \times 6\ k \cdot p$ model, which was developed for the strained wurtzite semiconductor to calculate the band structures.[19,20] The simulation also accounts the built-in polarization induced by the spontaneous and piezoelectric polarization and the effect of Shockley-Read-Hall (SRH) recombination.

AlGaN-based 284.5 nm DUV LED, which was fabricated by Yan et al.,[2] is used as a reference for the numerical simulation. The reference structure has a chip size of 400 μm x 400 μm, which is as depicted in Fig. 1. The doping densities which represent dopant concentration, are specified for each doped layer: $4\times10^{18} cm^{-3}$ for n-$Al_{0.6}Ga_{0.4}$N layer; $2\times10^{19} cm^{-3}$ for p-$Al_{0.65}Ga_{0.35}$N EBL; $2\times10^{19} cm^{-3}$ for p-$Al_{0.5}Ga_{0.5}$N interlayer; and $1\times10^{19} cm^{-3}$ for p-GaN layer. The number of QWs active region and quantum barriers (QBs) are varied, i.e. 2QBs/1QW; 3QBs/2QWs; 4QBs/3QWs; 5QBs/4QWs; and 6QBs/5QWs.

In the simulation, the value of the ionization energy of acceptors ($E_A$) is set to be 170 meV for p-GaN,[21] 470 meV for p-AlN[22], and scale linearly from 170 to 470 meV with Al composition for AlGaN.[3] A band offset ratio is set to be 0.65/0.35.[23] Other band structure parameters set to be similar with the recommended model for wurtzite nitride binaries at 300 K proposed by Vurgaftman et al[24], as summarized in Table 1. Polarization charges have crucial effect on the characteristics of nitride-based devices.[25] The Polarization charges induced by spontaneous and piezoelectric polarizations are computed by the interpolation methods developed by Fiorentini et al.[26] However, it has been reported that the experimental values of built-in polarization are weaker than theoretical values.[27, 28] It is mainly due to the fact that the built-in polarization is partly compensated by interface changes and structural defects.[29] In this work, the built-in polarization charges is assumed to be 50% of the theoretical value.[30] Moreover, Auger coefficient, SRH life time, and light extraction efficiency are set to be $5 \times 10^{-30}\ cm^6/s$, 10 ns, and 15%, respectively.[30-32]

Figure 2 shows the output power as a function of injection current for AlGaN/AlGaN DUV LEDs with 1-QW, 2-QWs, 3-QWs, 4-QWs, and 5-QWs. From Fig. 2, the 1-QW structure is advantaged when the current is lower than 24.75 mA, but becomes inferior to the 2-QWs, 3-QWs, 4-QWs, and 5-QWs when the current is higher than 27 mA, 31.25 mA, 34.5 mA, and 38.25 mA, respectively.

The results in Fig. 2 can be related to the results shown in Fig. 3, that the 1-QW structure has higher peak IQE but experience much bigger efficiency droop compared to the higher number of MQW structures, consequently it becomes lower than the IQE of MQW structures. In order to get a better idea between these two structures, the radiative recombination rate for 1-QW and 2-QWs are compared. Indeed, at 60 mA, the 2-QWs structure has a larger total radiative recombination rate than the 1-QW structure, as shown in Fig. 4. It agrees with the fact that at higher current, 2-QWs structure is superior than 1-QW structure. Thus, in the range of injection current under the study, 2-QWs structure is recommended for optimum AlGaN/AlGaN DUV LEDs, especially for the case of high current.

In summary, characteristics of AlGaN/AlGaN DUV LEDs with various number of QWs are numerically investigated. Simulation results show that 1-QW structure gives the best performance at very low injection current. However, at higher current, due to its high IQE and total recombination rate, the 2-QWs give the best performance among others. In overall, the 2-QWs structure is recommended to enhance the the device efficiency for AlGaN/AlGaN DUV LEDs.

**References**

1) M. A. Khan, M. Shatalov, H. P. Maruska, H. M. Wang, and E. Kuokstis, Jpn. J. Appl. Phys. **44**, 7191 (2005).

2) J. Yan, J. Wang, Y. Zhang, P. Cong, L. Sun, Y. Tian, C. Zhao, and J. Li, J. Cryst. Growth **414**, 254 (2015).

3) J.-Y. Chang, H.-T. Chang, Y.-H. Shih, F.-M. Chen, M.-F. Huang, and Y.-K. Kuo, IEEE Trans. Electron Dev., **64**, 12 (2017)

4) H. Hirayama, Y. Tsukada, T. Maeda, and N. Kamata, Appl. Phys. Exp. **3 (3)**, 031002 (2010).

5) A. Fujioka, T. Misaki, T. Murayama, Y. Narukawa, and T. Mukai, Appl. Phys. Exp. **3**, 4 (2010).

6) J. R. Grandusky, S. R. Gibb, M. C. Mendrick, C. Moe, M. Wraback, and L. J. Schowalter, Appl. Phys. Exp. **4**, 8 (2011).

7) Y. K. Kuo, F. M. Chen, J. Y. Chang, and Y. H. Shih, J. Appl. Phys. **119**, 094503 (2016).

8) S.-C. Huang, K.-C. Shen, D.-S. Wuu, P.-M. Tu, H.-C. Kuo, C.-C. Tu, R.-H. Horng, J. Appl. Phys. **110**, 12 (2011)

9) A. Knauer, V. Kueller, U. Zeimer, M. Weyers, C. Reich, and M. Kneissl, Phys. Status Solidi A **210**, 3 (2013).

10) M.-C. Tsai, S.-H. Yen, and Y.-K. Kuo, Appl. Phys. Lett. **98**, 11 (2011).

11) M.-F. Huang and T.-H. Lu, IEEE J. Quantum Electron. **42**, 8 (2006).

12) Y.-A. Chang, H.-C. Kuo, and T.-C. Lu., Semicon. Sci. Technol. **21**, 5 (2006).

13) J. Y. Chang, Y. K. Kuo, and M. C. Tsai, Phys. Status Solidi A **208**, 3 (2011).

14) T. Wang, D. Nakagawa, J. Wang, T. Sugahara, and S. Sakai, Appl. Phys. Lett. **73**, 3571 (1998).

15) S. Nakamura, M. Senoh, S. Nagahama, N. Iwasa, T. Matsushita, and T. Mukai, Appl. Phys. Lett. **76**, 22 (2000).

16) D.-J. Kim, Y.-T. Moon, K.-M. Song, C.-J. Choi, Y.-W. Ok, T.-Y. Seong, and S.-J. Park, J. Cryst. Growth **221**, 368 (2000).

17) Y.-K. Kuo, S.-H. Yen, and J.-R. Chen, *Nitride Semiconductor Devices* (Wiley, Berlin, 2007) p.294.

18) APSYS by Crosslight Software Inc., Burnaby, Canada (http://www. crosslight.com)

19) S.L. Chuang and C.S. Chang, Phys. Rev. B **54** (1996) 2491.

20) S.L. Chuang and C.S. Chang, Semicond. Sci. Technol. **12** (1997) 252.

21) W. Gotz, N. M. Johnson, J. Walker, D. P. Bour, and R. A. Street, Appl. Phys. Lett. **68**, 5

(1996).

22) J. Piprek, Appl. Phys. Lett. **104**, 5 (2014).

23) D. R. Hang, C. H. Chen, Y. F. Chen, H. X. Jiang, and J. Y. Lin, J. Appl. Phys. **90**, 4 (2001).

24) I. Vurgaftman and J. R. Meyer, J. Appl. Phys. **94**, 6 (2003).

25) E. F. Schubert, *Light-Emitting Diodes* (Cambridge, New York, 2006) 2nd Ed. p.224.

26) V. Fiorentini, F. Bernardini, and O. Ambacher, Appl. Phys. Lett. **90**, 4 (2002).

27) F. Renner, P. Kiesel, G. H. Dohler, M. Kneissl, C. G. Van de Walle, and N. M. Johnson, Appl. Phys. Lett. **81**, 3 (2002).

28) H. Zhang, E. J. Miller, E. T. Yu, C. Poblenz, and J. S. Speck, Appl. Phys. Lett. **84**, 23 (2004).

29) J. P. Ibbetson, P. T. Fini, K. D. Ness, S. P. DenBaars, J. S. Speck, and U. K. Mishra, Appl. Phys. Lett. **77**, 2 (2000).

30) Y.-H. Shih, J.-H. Chang, J.-K. Sheu, Y.-K. Kuo, F.-M. Chen, M.-L. Lee, W.-C. Lai, IEEE Trans. Electron Dev. **63**, 3 (2016).

31) H.-Y. Ryu, H.-S. Kim, and J.-I. Shim, Appl. Phys. Lett. **95**, 8 (2009).

32) J. Piprek and Z. S. Li, Appl. Phys. Lett. **102**, 2 (2013).

**Figure Captions**

**Fig. 1.** Schematic diagram of AlGaN/AlGaN deep UV LED

**Fig. 2**. Output power as a function of injection current for the AlGaN/AlGaN DUV LEDs with 1-QW, 2-QWs, 3-QWs, 4-QWs, and 5-QWs.

**Fig. 3.** Internal quantum efficiency as a function of injection current for the AlGaN/AlGaN DUV LEDs with 1-QW, 2-QWs, 3-QWs, 4-QWs, and 5-QWs.

**Fig. 4**. Radiative recombination rate for the AlGaN/AlGaN DUV LEDs with (a) 1-QW and (b) 2-QWs at 60 mA.

**Table I.** Band structure model for wurtzite nitride binaries at 300 K

| Parameters | Symbol (unit) | GaN | AlN | InN |
|---|---|---|---|---|
| Lattice constant | $a_0$ (Å) | 3.189 | 3.112 | 3.545 |
| Bandgap energy | $E_g$ (eV) | 3.42 | 6.0 | 0.64 |
| Spin-orbit splitting | $\Delta_{so}$ (eV) | 0.017 | 0.019 | 0.005 |
| Crystal-field splitting | $\Delta_{cr}$ (eV) | 0.010 | -0.169 | 0.040 |
| Hole effective mass | $A_1$ | -7.21 | -3.86 | -8.21 |
| | $A_2$ | -0.44 | -0.25 | -0.68 |
| | $A_3$ | 6.68 | 3.58 | 7.57 |
| | $A_4$ | -3.46 | -1.32 | -5.23 |
| | $A_5$ | -3.40 | -1.47 | -5.11 |
| | $A_6$ | -4.90 | -1.64 | -5.96 |
| Hydrost. deform. (c-axis) | $a_z$ (eV) | -4.9 | -3.4 | -3.5 |
| Hydrost. deform. (transverse) | $a_t$ (eV) | -11.3 | -11.8 | -3.5 |
| Shear deform. | $D_1$ (eV) | -3.7 | -17.1 | -3.7 |
| | $D_2$ (eV) | 4.5 | 7.9 | 4.5 |
| | $D_3$ (eV) | 8.2 | 8.8 | 8.2 |
| | $D_4$ (eV) | -4.1 | -3.9 | -4.1 |
| Elastic constant | $c_{33}$ (GPa) | 398 | 373 | 224 |
| | $c_{13}$ (GPa) | 106 | 108 | 92 |
| Piezoelectric coefficient | $d_{33}$ (pm/V) | 3.1 | 5.4 | 7.6 |
| | $d_{31}$ (pm/V) | -1.6 | -2.1 | -3.5 |
| Spontaneous polarization | $P_{sp}$ (C/m$^2$) | -0.034 | -0.09 | -0.042 |
| Electron effective mass (c-axis) | $m_e^z/m_o$ | 0.2 | 0.32 | 0.07 |
| Electron effective mass (transverse) | $m_e^t/m_o$ | 0.2 | 0.30 | 0.07 |

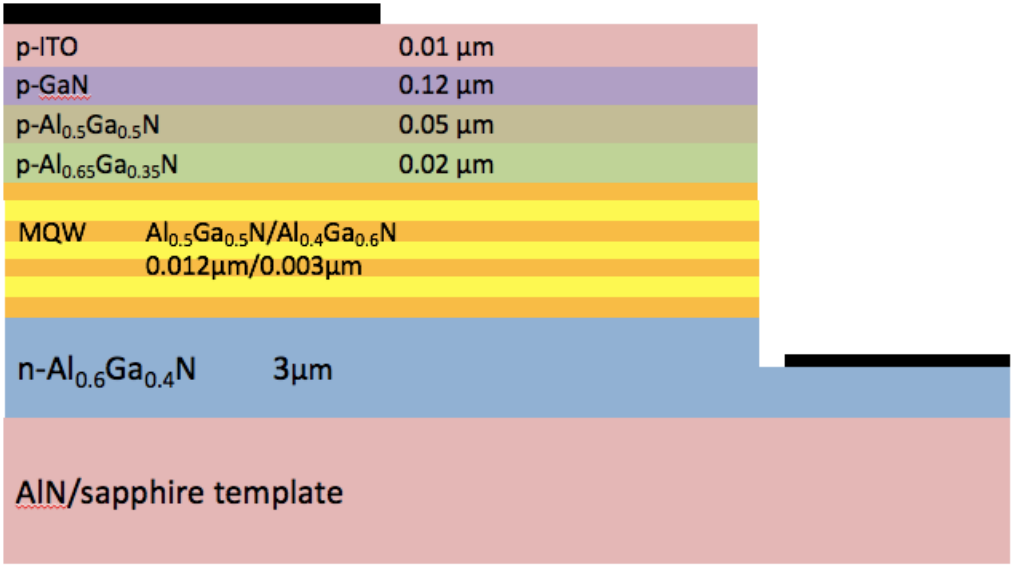

Fig.1.

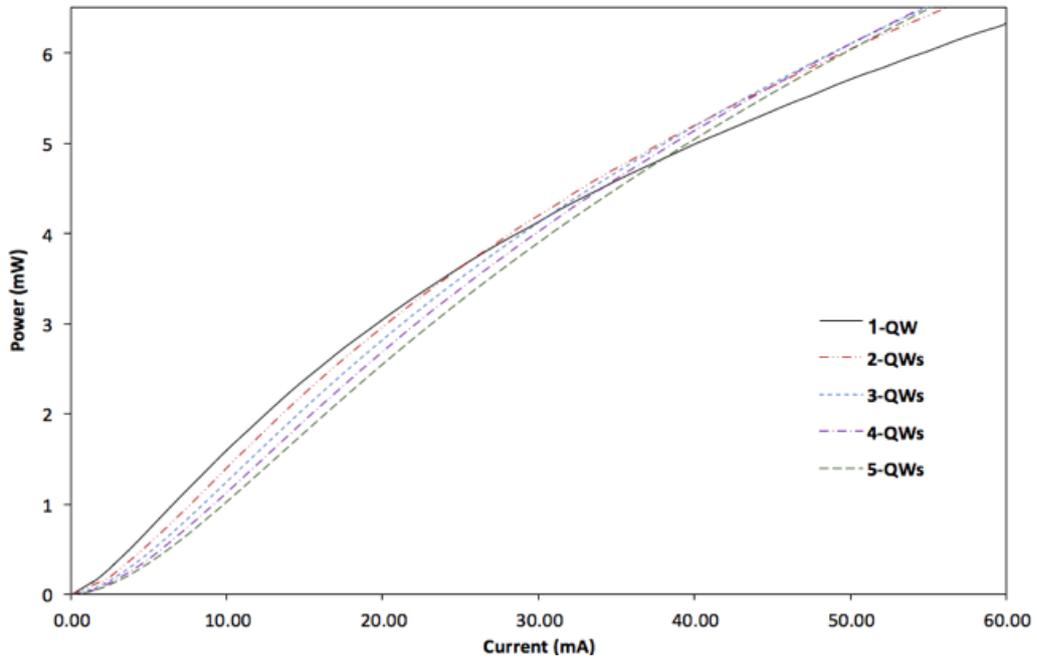

Fig. 2.

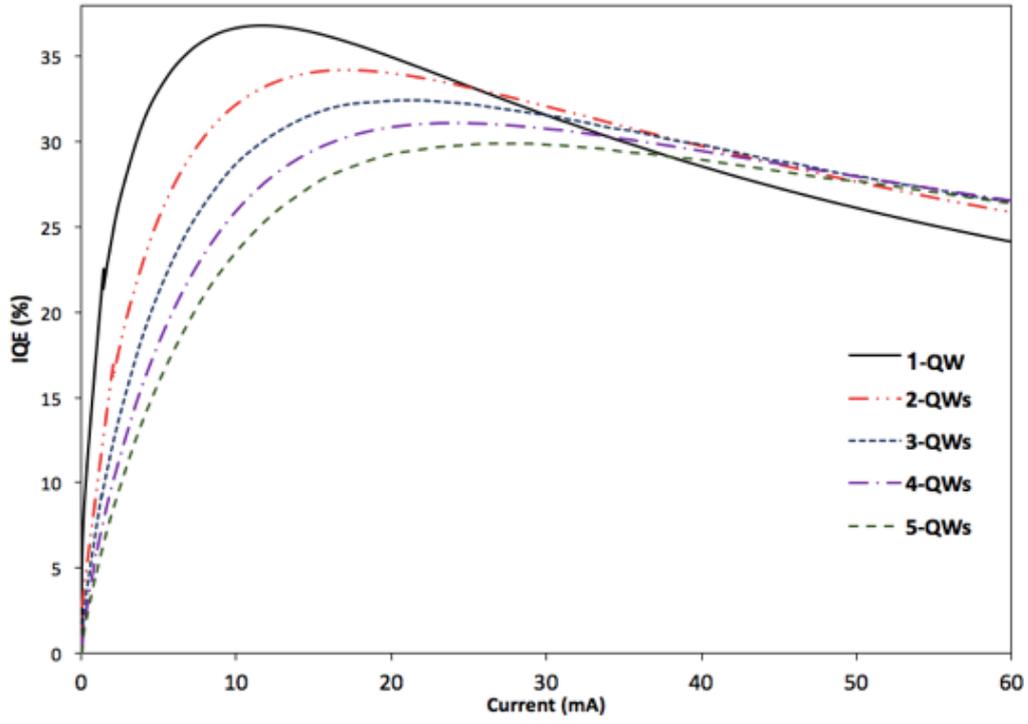

Fig. 3.

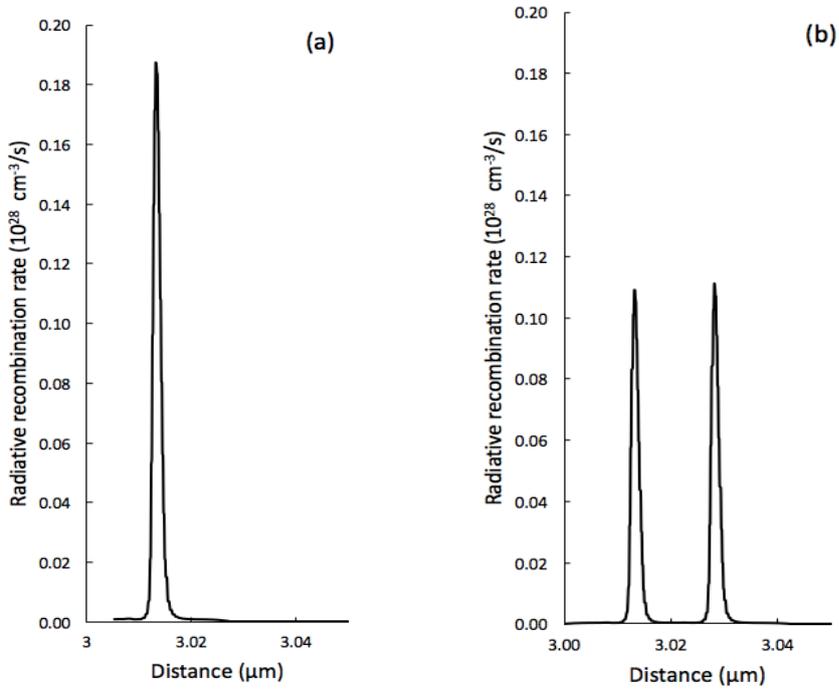

Fig. 4